\documentclass[12pt]{article}
\usepackage{amssymb,amsmath,graphics}
\def\R{\mathbb{R}}
\def\N{\mathbb{N}}
\def\Z{\mathbb{Z}}
\def\LL{\mathbb{L}}
\def\S{\mathbb{S}}
\def\T{\mathbb{T}}
\def\C{\mathbb{C}}

\def\demo{\noindent{\bf Proof:~~}}
\def\fin{\hfill$\square$\\}
\def\op#1{{\rm Op}\left(#1\right)}

\renewcommand{\Re}{\mathop{\rm Re}}
\renewcommand{\Im}{\mathop{\rm Im}}
\newtheorem{thm}{Theorem} [section]
\newtheorem{lem} [thm]{Lemma}
\newtheorem{prop}[thm]{Proposition}

\newtheorem{remark}[thm]{Remark}

\newenvironment{rem}{\begin{remark}\rm }{\end{remark}\medskip}
\makeatletter

\@addtoreset{equation}{section}
\makeatother
\begin{document}
\setcounter{footnote}{1}

\title{Scattering by a toroidal coil}
\author{ Ph. Roux\footnote{permanent address~: IUT Lannion, D\'epartement
Informatique, rue Branly BP150, 22302 Lannion CEDEX} \\
Department of Mathematics, University of Rennes,\\
 Campus Beaulieu, 35042, Rennes, France,\\
\tt Philippe.Roux@univ-rennes1.fr}
\date{}
\maketitle

\begin{abstract}
In this paper we consider the Schr\"odinger operator in $\R^3$ with a long-range magnetic potential associated to a magnetic field supported inside a torus $\T$. Using the scheme of smooth perturbations we construct stationary modified wave operators and the corresponding scattering matrix $S(\lambda )$. We prove that the essential spectrum of $S(\lambda )$ is an interval of the unit circle depending only on the magnetic flux $\phi$ across the section of $\T$. Additionally we show that, in contrast to the Aharonov-Bohm potential in $\R^2$, the total scattering cross-section is always finite. We also conjecture that the case treated here is a typical example in dimension 3.  
\end{abstract}


\section{Introduction}

Let $A(x)$ be a magnetic potential
\begin{equation}\label{eq:potA}
A(x)={a(\varphi_x)\over|x|}{\bf e}_{\varphi_x},~~|x|\geq R>0
\end{equation}
where $a\in C^\infty_0(0,\pi)$ is a positive function of the colatitude of $x=(x_1,x_2,x_3)$
\begin{equation}\label{eq:colat}
\varphi_x=\arccos\left({x_3\over|x|}\right)\in[0,\pi],
\end{equation}
and ${\bf e}_{\varphi_x}$ denotes the unitary vector of spherical coordinates
$\left({\bf e}_{r_x} ,{\bf e}_{\varphi_x},{\bf e}_{\theta_x} \right)$ associated to the point $x$
\begin{equation}\label{eq:ephi}
{\bf e}_{\varphi_x}={1\over|x|}\left(
{x_1x_3\over \sqrt{x_1^2+x_2^2}},
{x_2x_3\over \sqrt{x_1^2+x_2^2}},
-\sqrt{x_1^2+x_2^2}\right).
\end{equation}
Physically the potential (\ref{eq:potA}) corresponds to a magnetic field $B={\rm curl} A$ supported inside a torus $\T$ obtained by revolution around the $x_3$ axis (see Fig. \ref{fig:solenoid}). The function $a$ in (\ref{eq:potA}) depends only on the section of $\T$ and the flux of $B$ across this section is
\begin{equation}\label{eq:flux}
\phi=\int_0^\pi a(\varphi)d\varphi>0.
\end{equation}
This situation is kwown, from the work of Aharonov and Bohm \cite{AB1}, to show a purely quantum phenomena~: a compactly supported magnetic field can act on particles which never cross its support. From the mathematical point of view, despite $B$ as a finite support, the potential $A$ decays as $|x|^{-1}$ at infinity and is of long-range nature. Thus one expects that the properties of the scattering process associated to the potential (\ref{eq:potA}) be different from the case of short-range potentials. \\

Here we consider the Schr\"odinger operator
\begin{equation}\label{eq:H}
H=(D-A(x))^2,~~D=-i\nabla_x,
\end{equation}
in $\LL^2(\R^3)$, and develop the scheme of smooth perturbations for the pair $H,H_0=-\Delta$.
Although usual wave operators exists (due to the transversal gauge condition $<A(x),x>=0,$ for all $x\in\R^3$, see \cite{LOSTHA}) we prefer to work with modified wave operators of the Isozaki-Kitada type~:
\begin{equation}\label{eq:MWO}
W_\pm(H,H_0,J)=s-\lim_{t\to\pm\infty}e^{itH}Je^{-itH_0}
\end{equation}
with stationary identifications $J=J_\pm$ depending on the sign of $t$, as in \cite{NICO1,ROYA}. We choose the operators $J_\pm$ as pseudo-differential operators (PDO) with symbols $\exp({i\Phi_\pm(x,\xi)})$ such that the effective perturbation $T_\pm=HJ_\pm-J_\pm H_0$ be short-range, that is the phase-function $\Phi_\pm$ satisfies $\nabla_x\Phi_\pm(x,\xi)=A(x)$. Thus the existence of wave operators (\ref{eq:MWO}) relies only on the \emph{limiting absorption principle} in contrast to \cite{ROYA} where the \emph{radiation estimate} was needed. Since the identifications $J_\pm$ are ``close'' to unitary operators, the wave operators $W_\pm(H,H_0,J_\pm)$ are automatically isometric and complete, indeed, they coincide with the usual wave operators $W_\pm(H,H_0)=W_\pm(H,H_0,Id)$.\\

The scattering operator, defined by $S=W_+^*(H,H_0)W_-(H,H_0)$, commutes with $H_0$, so, in the spectral representation of $H_0$, it reduces to the multiplication by the operator valued function $S(\lambda)$, named the scattering matrix (SM) which acts as an integral operator on the unit sphere $\S^2$ of $\R^3$. Our study of the SM relies on its stationary representation
\begin{equation}\label{eq:SM}
S(\lambda)={\cal W}(\lambda)-2i\pi\Gamma_0(\lambda)\left(J_+^*T_--T_+^*R(\lambda+i0)T_-\right)\Gamma_0^*(\lambda)
\end{equation}
where $R(z)=(H-z)^{-1}$, and
\begin{equation}\label{eq:SAB}
{\cal W}(\lambda)=\Gamma_0(\lambda)W_+(H_0,H_0,J_+^*J_-)\Gamma_0^*(\lambda)
\end{equation}
with $\Gamma_0(\lambda):\LL^2(\R^3)\longrightarrow \LL^2(\S^2)$ defined for $u$ in the Schwarz class by
\begin{equation}\label{eq:Gam0}
\left(\Gamma_0(\lambda)u\right)(\omega)={\sqrt{\lambda}\over 2(2\pi)^{3/2}}\int_{\R^3}e^{i\sqrt{\lambda}<\omega,x>}u(x) dx,~~\omega\in\S^2,
\end{equation}
and $\Gamma_0^*(\lambda)$ is the formally adjoint to $\Gamma_0(\lambda)$. To justify the formula (\ref{eq:SM}) we decompose it as a sum of bounded operators. First we calculate the term ${\cal W}(\lambda)$ and prove that it reduces to the operator of multiplication by the function ${\rm w}$ defined on $\S^2$ by
\begin{equation}\label{eq:wplus}
{\rm w}(\omega)=\exp\left(i\int^{\varphi_\omega}_{\pi-\varphi_\omega}a(\varphi)d\varphi\right),~~\omega\in\S^2.
\end{equation}
Then we show that the remaining term, $S(\lambda)-{\cal W}(\lambda)$, is an integral operator on $\S^2$ with a $C^\infty$ kernel. Thus, we can make a spectral analysis of the SM. Since $S(\lambda)$ is a compact perturbation of ${\cal W}(\lambda)$, we calculate its essential spectrum, that is
\begin{equation}\label{eq:spec}
\sigma_{ess}\left(S(\lambda)\right)=\left\{\mu=\exp(i\nu)\in\C~|~\nu\in[-\phi,\phi]\right\}.
\end{equation}
In particular $\sigma_{ess}(\lambda)$ depends only on the magnetic flux $\phi$ (\ref{eq:flux}) of $B$ across the section of $\T$. Now if we take as a definition of the differential scattering cross-section
\begin{equation}\label{eq:diffscat}
\Sigma_{diff}(\omega,\omega_0;\lambda)=
{\lambda^{(-d-1)/2}\over (2\pi)^{d-1}}
|{\rm s}(\omega,\omega_0;\lambda)|^2,
~~\omega\neq\omega_0,
\end{equation}
with $d=3$ and where $\omega_0$(resp. $\omega$) is the in-coming (out-going) direction, then the function $\Sigma_{diff}(\omega,\omega_0;\lambda)$ belongs to $C^\infty(\S^2\times\S^2\times\R^+)$. In particular the total scattering cross-section
\begin{equation}\label{eq:totalscat}
\Sigma_{tot}(\omega_0;\lambda)=\int_{\S^2}
\Sigma_{diff}(\omega,\omega_0;\lambda)d\omega
\end{equation}
is finite for all incident directions $\omega_0\in\S^2$.\\

The paper is organized as follow~: in section \ref{WO} we construct stationary wave operators and recover the basic results of scattering theory for potential (\ref{eq:potA}); in section \ref{SMp} we analyze the structure of the SM and its spectral properties; finally, in section \ref{ABp} we make some remarks about this example and the 2-dimensional Aharonov-Bohm effect, we also conjecture that the situation described here is very general in the 3-dimensional case.


\section{Wave operators}\label{WO}

In this section we construct time-independant modified wave operators, as in \cite{NICO1,ROYA}, and recover basic results on long-range magnetic scattering in transversal gauge \cite{LOSTHA}.\\

\begin{figure}[h]
\begin{center}
\includegraphics{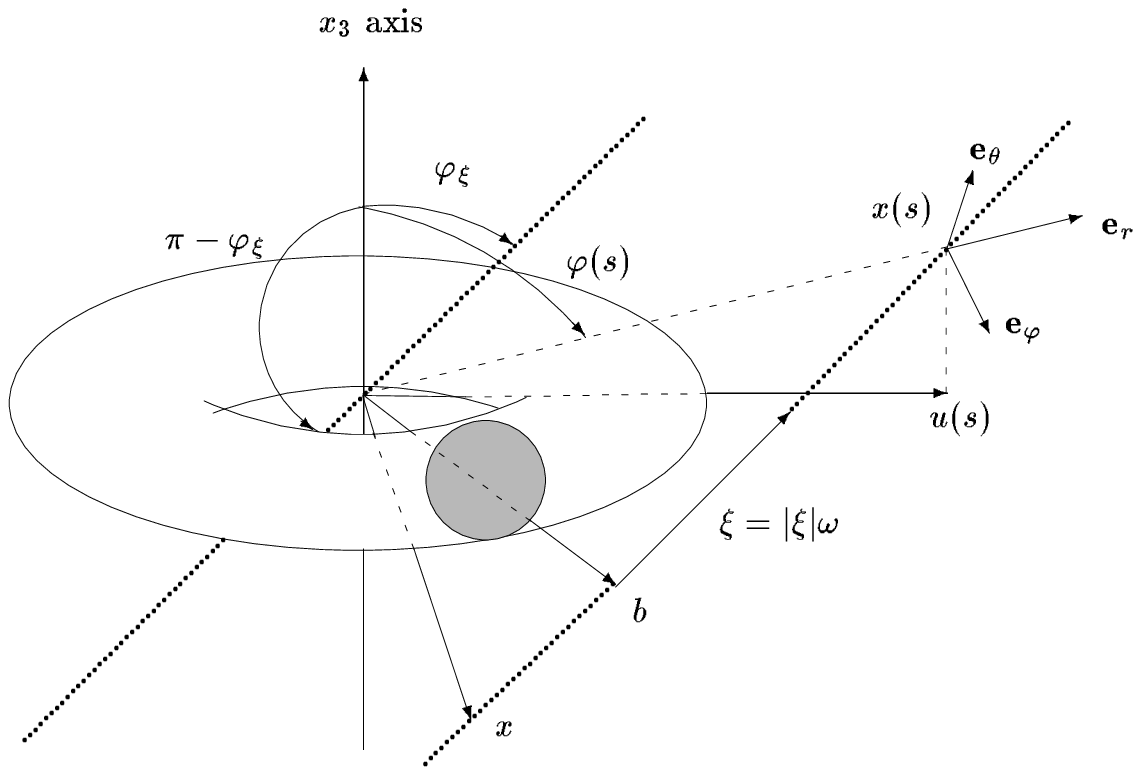}
\caption{the toroidal coil $\T$}\label{fig:solenoid}
\end{center}
\end{figure}

{\bf 2.1.} In the scheme of smooth perturbations the choice of identifications $J=J_\pm$ in (\ref{eq:MWO}) is determined by the condition that the effective perturbation $T_\pm=HJ_\pm-J_\pm H_0$ be ``short-range''. If, as in \cite{YAF6}, we search $J_\pm$ as a PDO with symbol $j_\pm(x,\xi)$ then the function $\Psi_\pm(x,\xi)=e^{i<x,\xi>}j_\pm(x,\xi)$ should be an approximate ({\it i.e.} up to short-range terms) eigenfunction of $H$ assiociated to the eigenvalue $|\xi|^2$. Thus we set $j_\pm(x,\xi)=\exp(i\Phi_\pm(x,\xi))$ and compute
\begin{eqnarray}(H-|\xi|^2)\Psi_\pm(x,\xi)
&=&\Big(2<\xi,\nabla_x\Phi_\pm(x,\xi)-A(x)>+
|\nabla_x\Phi_\pm(x,\xi)-A(x)|^2\nonumber \\
&&~~-i{\rm div}_x\left(\nabla_x\Phi_\pm(x,\xi)-A(x)\right)\Big)\Psi_\pm(x,\xi).\label{eq:eik}
\end{eqnarray}
Taking only the principal ({\it i.e.} the first) term of (\ref{eq:eik}), we obtain the eikonal equation for $\Phi_\pm$
\begin{equation}\label{eq:eikonal}
<\xi,\nabla_x\Phi_\pm(x,\xi)-A(x)>=0.
\end{equation}
As shown in \cite{YAF6}, this equation admits solutions with decaying derivatives for large $\vert x\vert$
$$\Phi_\pm(x,\xi)=\mp\int_0^\infty<A(x\pm t\xi)-A(\pm t\xi),\xi>dt=\mp\int_0^\infty<A(x\pm t\xi),\xi>dt.$$
Note that the second equality is a consequence of the transversal gauge condition 
$<A(y),y>=0$, for all $y\in\R^3$.
 To simplify this expression we first make the change of variables $t\mapsto s$ defined by
$$s=s_0\pm t|\xi|,~~x=b+s_0\omega,~~<b,\omega>=0,~~\omega={\xi\over|\xi|},$$
which leads to the equation
\begin{equation}\label{eq:circ}
\Phi_\pm(x,\xi)=\int^{s_0}_{\pm\infty}<A(b+ s\omega),\omega>ds.
\end{equation}
Then, we rewrite this integral into spherical coordinates (see Fig. \ref{fig:solenoid}). Let
$$x(s)=b+s\omega,~~u(s)=\sqrt{(b_1+s\omega_1)^2+(b_2+s\omega_2)^2},$$
and $\varphi(s)$ be the colatitude of $x(s)$ (defined by (\ref{eq:ephi})). Since
$$\sin(\varphi(s))=u(s)/|x(s)|,~~\cos(\varphi(s))=(b_3+s\omega_3)/|x(s)|,$$
and taking into account that
$$ |x|^2=|b|^2+s^2,~|\omega|^2=1,~<x(s),\omega>=s,~{d\over ds}\cos(\varphi(s))=-\sin(\varphi(s)){d\varphi(s)\over ds},$$
we get
$${d\varphi(s)\over ds}={sb_3-\omega_3|b|^2\over |x(s)|^2 u(s)}.$$
On the other hand
\begin{eqnarray*}
<{\bf e}_{\varphi(s)},\omega>
&=&{1\over|x(s)|u(s)}\left(x_3(s)<x(s),\omega>-<(0,0,|x(s)|^2),\omega>\right)\\
&=&{|x(s)|^2\omega_3-sx_3(s)\over|x(s)|u(s)}={sb_3-|b|^2\omega_3 \over|x(s)|u(s)}
\end{eqnarray*}
which leads to
\begin{eqnarray*}
<A(x(s)),\omega>&=&{a(\varphi(s))\over |x(s)|}<{\bf e}_{\varphi(s)},\omega>\\
&=&{a(\varphi(s))\over |x(s)|}{sb_3-|b|^2\omega_3\over|x(s)|u(s)}=
{a(\varphi(s))}{d\varphi(s)\over ds}.
\end{eqnarray*}
Thus, we can make the change of variables $s\mapsto\varphi(s)$ in (\ref{eq:circ}) and, since $\varphi(s_0)=\varphi_x$ and $\varphi(\pm\infty)=\varphi_{\pm\omega}=\varphi_{\pm\xi}$, we get
\begin{equation}\label{eq:Phi}
\Phi_\pm(x,\xi)=\int_{\varphi_{\pm\xi}}^{\varphi_x}a(\varphi)d\varphi,~~|x|\geq R>0,
\end{equation}
in particular, for $|x|\geq R>0$,
\begin{equation}\label{eq:PhiA}
\nabla_x\Phi_\pm(x,\xi)={1\over |x|}\partial_{\varphi_x}\Phi_\pm(x,\xi){\bf e}_{\varphi_x}=A(x).
\end{equation}

The stationary scheme developped below make an intensive use of symbolic calculus (see \cite{TAY}), so we have to fix some notations on PDO. In the following we call ${\cal S}^m(\mu)$ the set of functions $p\in C^\infty(\R^{6})$ satisfying, for all multi-indices $\alpha$ and $\beta$, the estimates
$$|\partial_x^\alpha\partial_\xi^\beta p(x,\xi)|\leq C_{\alpha,\beta}\left<x\right>^{m-|\alpha|}\left<\xi\right>^{\mu-|\beta|}$$
and ${\cal S}^m=\cap_{\mu\in\Z}{\cal S}^m(\mu)$. We set $P=\op{p(x,\xi)}=p(x,D)$ for the PDO with symbol $p\in{\cal S}^m$ defined for $u\in{\cal S}(\R^3)$ by
$$(Pu)(x)=\int_{\R^3}e^{i<x,\xi>}p(x,\xi)\widehat{u}(\xi){d\xi\over(2\pi)^{3/2}},$$
where $\widehat{u}$ denotes the Fourier transform of $u$
\begin{equation}\label{eq:fourier}
\widehat{u}(\xi)=\int_{\R^3}e^{-i<x,\xi>}u(x){dx\over(2\pi)^{3/2}}.
\end{equation}
With this notations an operator with symbol $p\in{\cal S}^m$ is bounded (compact) if $m\leq0$ $(m<0)$. Now we are able to define the identifications $J_\pm$.
\begin{lem}\label{Jpm}
Let us fix $\lambda\in(0,+\infty)$, $r\in(0,\lambda/2)$, and let $\psi,\eta\in C^\infty(\R^3,[0,1])$ be cut-off functions satifying~:
\begin{description}
\item[i)] $\eta(x)=0$ if $|x|\leq R$ and $\eta(x)=1$ if $|x|\geq R+1$,
\item[ii)] $\psi(\xi)=1$ if $||\xi|-\lambda|\leq r$ and $\psi(\xi)=0$ if $||\xi|-\lambda|\geq 2r$.
\end{description}
For any choice of functions $\eta$ and $\psi$ we set $J_\pm=\op{j_\pm(x,\xi)}$ where
\begin{equation}\label{eq:Jpm}
j_\pm(x,\xi)=e^{i\Phi_\pm(x,\xi)}\eta(x)\psi(\xi).
\end{equation}
Then $j_\pm\in {\cal S}^0$ and $J_\pm$ is a bounded operator in $\LL^2(\R^3)$. Additionally the symbol of the effective perturbation $T_\pm=HJ_\pm-J_\pm H_0=\op{t_\pm(x,\xi)}$ belongs to ${\cal S}^m$ for all $m\in\Z$.
\end{lem}
\demo Since, by (\ref{eq:Phi}), $\Phi_\pm(x,\xi)$ is an homogeneous function of $x$ and $\xi$ (for $|x|\geq R$) of degree $0$, it satisfies in the whole phase-space  the estimate~:
$$|\partial_x^\alpha\partial_\xi^\beta \Phi_\pm(x,\xi)|\leq C_{\alpha,\beta}|x|^{-|\alpha|}|\xi|^{-|\beta|},~~\forall \alpha,\beta\in\N^3,~~|x|\geq R.$$
Thus taking into account the definition of cut-off functions $\eta$ and $\psi$ we get that $j_\pm\in{\cal S}^0$ and $J_\pm$ is bounded by the Calderon-Vaillancourt Theorem. Now we calculate the symbol of the effective perturbation
$$t_\pm(x,\xi)=e^{-i<x,\xi>}(H-|\xi|^2)e^{i<x,\xi>}j_\pm(x,\xi).$$
Using (\ref{eq:eik}), (\ref{eq:Jpm}), and that $\nabla_x\Phi_\pm(x,\xi)=A(x)$, by (\ref{eq:PhiA}), we get
$$t_\pm(x,\xi)=e^{i\Phi_\pm(x,\xi)}\left(-2i<\xi,\nabla_x\eta(x)>-\Delta_x\eta(x)\right)\psi(\xi).
$$
Thus $t_\pm$ has a compact support in $x$ and $\xi$ and belongs to ${\cal S}^m$ for all $m\in\Z$.\fin

{\bf 2.2.} Our proof of the existence and asymptotic completeness of modified wave operators (\ref{eq:MWO}) is based on the scheme of smooth perturbations. Then it relies on the well-known limiting absorption principle~:

\begin{thm}[Limiting Absorption Principle]\label{LAP}
Let $H$ be the operator $(\ref{eq:H})$ with potential $(\ref{eq:potA})$. Then, for all bounded interval $\Lambda\subset(0,\infty)$, disjoint from $0$, the operator function
$\left<x\right>^{-s}R(z)^n\left<x\right>^{-s},$ $s>n-1/2$, is $($H\"older-$)$
continuous in norm in the region $\Re(z)\in\Lambda,\pm\Im(z)\in(0,1]$ and

\begin{equation}\label{eqLAP1}
\mathop{\sup_{\Re z\in\Lambda}}_{1\ge|\Im z|>0}
\|\left<x\right>^{-s}R(z)^n\left<x\right>^{-s}\|\leq c, ~~
\forall s>n-1/2.
\end{equation}
In particular the spectrum of $H$ in $\Lambda$ is absolutely continuous
and the operators $\left< x\right>^{-s},~s>1/2$ are $H$-smooth on
$\Lambda$ $($in the sense of Kato$)$.
\end{thm}

This result can easily be derived from the Mourre commutator method \cite{JEN} and the absence of positive eigenvalues for the operator (\ref{eq:H}) \cite{IKUC}. Now, since $t_\pm$ is short-range in the whole space, our proof of existence and completeness of wave operators relies only on the theorem \ref{LAP} in contrast to \cite{ROYA} where the radiation estimate was also needed.

\begin{prop}\label{MWO}
Let $E$ and $E_0$ be, respectively, the spectral mesure of $H$ and $H_0$, and $J_\pm$ be constructed as under the assumptions $\rm i)$ and $\rm ii)$ of lemma $\ref{Jpm}$
Set $\Lambda=(\lambda-r,\lambda+r)$ then the wave operators $W_\pm(H,H_0,J_\pm)$ and $W_\pm(H_0,H,J_\pm^*)$ exist, are isometric on, respectively, $E_0(\Lambda)$ and $E(\Lambda)$ and are adjoint one to each other. Additionally asymptotic completeness holds for the triple $(H,H_0,J_\pm)$, that is
\begin{eqnarray*}
{\rm Ran}\left( W_\pm(H,H_0,J_\pm)E_0(\Lambda)\right)&=&{\rm Ran}\left( E(\Lambda)\right),\\
{\rm Ran}\left( W_\pm(H_0,H,J_\pm^*)E(\Lambda)\right)&=&{\rm Ran}\left( E_0(\Lambda)\right).
\end{eqnarray*}
\end{prop}
\demo
Since the operators $\left<x\right>^{-s}$ are $H$ and $H_0$-smooth for all $s>1/2$, the effective perturbation admits a decomposition into a product of smooth perturbations
\begin{equation}\label{eq:fact}
T_\pm=\left<x\right>^{-1}
\left(\left<x\right>T_\pm\left<x\right>\right)
\left<x\right>^{-1}
\end{equation}
because the PDO $\left<x\right>T_\pm\left<x\right>$ belong to ${\cal S}^m$ for all $m\in\Z$ and so is a bounded operator. This is sufficient to prove the existence of $W_\pm(H,H_0,J_\pm)$ and $W_\pm(H_0,H,J_\pm^*)$ which are obviously adjoint one to each other. Now, by the chain rule, isometricity and completeness of $W_\pm(H,H_0,J_\pm)$ are, respectively, equivalent to
\begin{eqnarray*}
W_\pm(H_0,H,J_\pm^*) W_\pm(H,H_0,J_\pm)E_0(\Lambda)&=&W_\pm(H_0,H_0,J_\pm^*J_\pm)E_0(\Lambda)=E_0(\Lambda),\\
W_\pm(H,H_0,J_\pm) W_\pm(H_0,H,J_\pm^*)E(\Lambda)&=&W_\pm(H,H,J_\pm J_\pm^*)E(\Lambda)=E(\Lambda).
\end{eqnarray*}
The operator $J_\pm^* J_\pm-\psi^2(D)$ is compact since its principal symbol, equals to $(\eta(x)-1)\psi^2(\xi)$, is compactly supported in $x$. Together with the identity 
$\psi^2(D)E_0(\Lambda)=E_0(\Lambda)$ this leads to 
\begin{eqnarray*}
W_\pm(H_0,H_0,J_\pm^*J_\pm)E_0(\Lambda)&=&W_\pm(H_0,H_0,\psi^2(D)E_0(\Lambda))\\
&=&W_\pm(H_0,H_0,E_0(\Lambda))=E_0(\Lambda).
\end{eqnarray*}
Then, $W_\pm(H,H_0,J_\pm)$ are isometric. Asymptotic completeness goes on the same way remarking also that $E(\Lambda)-E_0(\Lambda)$ is compact.\fin

Finally let us check that the wave operators constructed here coincide with the usual ones constructed in \cite{LOSTHA}.

\begin{prop}\label{UWO}
Under assumptions $\rm i)$ and $\rm ii)$ of lemma $\ref{Jpm}$ we have 
$$W_\pm(H,H_0,J_\pm)=W_\pm(H,H_0,Id)\psi(D)$$
for any choice of functions $\eta$ and $\psi$.
\end{prop}

\demo The proof relies on the stationary phase formula applied to the integral
$$\left(J_\pm e^{-itH_0}u\right)(x)=
\int_{\R^3}e^{i<x,\xi>-it|\xi|^2}j_\pm(x,\xi)\widehat{u}(\xi){d\xi\over(2\pi)^{3/2}}.$$
Since the stationary points $\xi_0=\xi_0(t)$ of this integral are $\xi_0=x/(2t)$ we get, for $u\in{\cal S}(\R^3)$, the asymptotics
$$\left(J_\pm e^{-itH_0}u\right)(x)={e^{\mp id\pi/4}\over (2t)^{d/2}}e^{i|x|^2/(2t)+i\Phi_\pm(x,x/(2t))}\widehat{u}(x/(2t))\eta(x)\psi(x/(2t))+{\rm r}_\pm(x,t)$$
where ${\rm r}_\pm(x,t)$ tends to $0$ in $\LL^2(\R^3)$ as $t\to\pm\infty$. Now, by (\ref{eq:Phi}), we have $\Phi_\pm(x,x/(2t))=0$ for $|x|\geq R$ so the phase factor $\exp(i\Phi_\pm(x,x/(2t)))$ is inessential and
$$\lim_{t\to\pm\infty}\left(J_\pm e^{-itH_0}- e^{-itH_0}\right)u
=\lim_{t\to\pm\infty}(\eta-1)e^{-itH_0}\psi(D)u=0$$
since $\eta-1$ is $H_0$-compact. In conclusion usual wave operators exists and co\"\i ncide with the wave operators of proposition \ref{MWO}.\fin

\begin{rem}
If a ``short-range" electromagnetic perturbation $(V_0,A_0)$
\begin{equation}\label{eq:shortR}
|V_0(x)| +|A_0(x)|+|{\rm div}\, A_0(x)|\leq C\left<x\right>^{-\rho},~~\rho>1,
\end{equation}
is added to the operator $H$, then all the results of this section remain true without changing the definition (\ref{eq:Jpm}) of identifications $J_\pm$. The additional term $\widetilde{T}_\pm$ arising in the effective perturbation $T_\pm$ are short-range. Since $\rho>1$, and by theorem \ref{LAP}, they admit a factorization into a product of $H$-smooth operators similar to (\ref{eq:fact})
$$\widetilde{T}_\pm=\left<x\right>^{-\rho/2}\left(\left<x\right>^{\rho/2}
\widetilde{T}_\pm\left<x\right>^{\rho/2}\right)\left<x\right>^{-\rho/2}.$$
\end{rem}


\section{The scattering matrix}\label{SMp}

In this section we consider the SM for the pair $H,H_0$ and its stationary representation. We do not give a proof of formula (\ref{eq:SM}) (a complete justification can be found in \cite{YAF8}), but we rewrite it into a sum of bounded operators on $\LL^2(\S^2)$ which gives its precise meaning to the formula (\ref{eq:SM}). Thus we can make the analysis of spectral properties and singularities of $S(\lambda)$ for all $\lambda>0$.\\

Let us decompose formula (\ref{eq:SM}) as follow
\begin{eqnarray}
\label{eq:S}S(\lambda)&=&{\cal W}(\lambda)+S_1(\lambda)+S_2(\lambda)\\
\label{eq:S1}S_1(\lambda)&=&-2i\pi\Gamma_0(\lambda)J_+^*T_-\Gamma_0^*(\lambda)\\
\label{eq:S2}S_2(\lambda)&=&2i\pi\Gamma_0(\lambda)T_+^*R(\lambda+i0)T_-\Gamma_0^*(\lambda)
\end{eqnarray}
with ${\cal W}(\lambda)$ and $\Gamma_0(\lambda)$ given by (\ref{eq:SAB}) and (\ref{eq:Gam0}). In the following three propositions we analyze separately the terms ${\cal W}(\lambda),S_1(\lambda)$ and $S_2(\lambda)$.

\begin{prop}\label{SAB}
The operator ${\cal W}(\lambda)$ defined by $(\ref{eq:SAB})$ is the operator of multiplication by the function ${\rm w}(\omega)$ defined on $\S^2$ by $(\ref{eq:wplus})$.
\end{prop}
\demo First remark that the commutator 
$$[H_0,J_+^*J_-]=T_+^*J_-+J_+^*T_-$$ 
admits  a factorization into a sum of products of $H_0$-smooth operators. Then the wave operator $W_+(H_0,H_0,J_+^*J_-)$ is well defined, commutes with $H_0$ (by the interwinning property), and so it reduces to multiplication by the operator valued function ${\cal W}(\lambda)$ in the spectral representation of $H_0$. Up to compact terms the operator $J_+^*J_-$ is the PDO with principal symbol $\exp(i\Theta(x,\xi))$ with
$$\Theta(x,\xi) =\Phi_-(x,\xi)-\Phi_+(x,\xi),$$
taking into account (\ref{eq:Phi}) we obtain that $\Theta$ does not depends on $x$
\begin{equation}\label{eq:Theta}
\Theta(x,\xi)=\int^{\varphi_x}_{\varphi_{-\xi}}a(\varphi)d\varphi-\int^{\varphi_x}_{\varphi_{\xi}}a(\varphi)d\varphi=\int^{\varphi_\xi}_{\varphi_{-\xi}}a(\varphi)d\varphi=:\Theta(\xi).
\end{equation}
Now since the operator $\exp(i\Theta(D))$ commutes with $H_0$ we get
\begin{eqnarray*}
W_+(H_0,H_0,J_+^*J_-)E_0(\Lambda)&=&
s-\lim_{t\to\pm\infty}e^{-itH_0}J_+^*J_-e^{-itH_0}E_0(\Lambda)\\
&=&s-\lim_{t\to\pm\infty}e^{-itH_0}e^{i\Theta(D)}e^{-itH_0}E_0(\Lambda)\\
&=&e^{i\Theta(D)}E_0(\Lambda).
\end{eqnarray*}
The function $\Theta$ is obviously homogeneous of degree $0$, by (\ref{eq:Theta}), Together with the obvious identity $\varphi_{-\omega}=\pi-\varphi_\omega$  this leads to $\exp(i\Theta(\sqrt{\lambda}\omega))=\exp(i\Theta(\omega))={\rm w}(\omega)$. Then in the spectral representation where $H_0$ is diagonal the operator ${\cal W}(\lambda)$ reduces to the operator of multiplication by the function (\ref{eq:wplus}).
\fin

\begin{rem}\label{phys}
From the physical point of view, $\Theta(x,\xi)$ is the circulation of the magnetic potential $A(x)$ along the ``closed" contour symbolized by doted lines on Figure \ref{fig:solenoid}. In particular the calculation of function $\Theta$ is independent of the gauge choosen for $A(x)$. Thus the scheme developed before applies to any magnetic potential $\tilde{A}(x)$ satisfying ${\rm curl}(\tilde{A})={\rm curl}(A)$, however usual wave operators $W_\pm(\tilde{H},H_0)$, with $\tilde{H}=(D-\tilde{A})^2$, should not exists if the transversal gauge is not assumed.
\end{rem}

Here we note that the kernel of ${\cal W}(\lambda)$ is ${\rm w}(\omega)\delta(\omega,\omega')$ where $\delta$ denotes the Dirac distribution on $\S^2$. Below we show that the kernel of $S(\lambda)$ does not contains any other singularity.\\

\begin{prop}\label{S1}
The operator $S_1(\lambda)$, defined in $(\ref{eq:S1})$, is an integral operator on $\S^2$ with a smooth kernel ${\rm s}_1(\omega,\omega';\lambda)\in C^\infty(\S^2\times\S^2\times\R^*_+)$. In particular $S_1(\lambda)$ belongs to the Hilbert-Schmidt class.
\end{prop}

\demo
By equations (\ref{eq:Gam0}), (\ref{eq:fourier}), the operator $S_1(\lambda)=-2i\pi\Gamma_0(\lambda)J_+^*T_-\Gamma_0^*(\lambda)$ is the restriction of a PDO on $\LL^2(\R^3_\xi)$ with amplitude $\overline{j_+(x,\xi)}t_-(x,\xi')$ to the sphere $|\xi|^2=|\xi'|^2=\lambda$, thus it is an integral operator on $\LL^2(\S^2)$ with kernel
\begin{equation}\label{eq:kers1}
{\rm s}_1(\omega,\omega';\lambda)=-{i\sqrt{\lambda}\over8\pi^2} \int_{\R^3}e^{i\sqrt{\lambda}<\omega'-\omega,x>}\overline{j_+(x,\sqrt{\lambda}\omega)}t_-(x,\sqrt{\lambda}\omega')dx.
\end{equation}
Since the amplitude $\overline{j_+(x,\sqrt{\lambda}\omega)}t_-(x,\sqrt{\lambda}\omega')$ is compactly supported in $x$ (due to the presence of derivatives of function $\eta$ defined in lemma \ref{Jpm}) the integral above obviously converge. Differentiating expression (\ref{eq:kers1}) we get that ${\rm s}_1(\omega,\omega';\lambda)$ is a $C^\infty$-function. In particular $|{\rm s}_1(\omega,\omega';\lambda)|^2$ is bounded and the Hilbert-Schmidt norm
$$\int_{\S^2}\int_{\S^2}|{\rm s}_1(\omega,\omega';\lambda)|^2d\omega d\omega'$$
of $S_1(\lambda)$ is finite.\fin

\begin{prop}\label{S2}
The operator $S_2(\lambda)$, defined in $(\ref{eq:S2})$, is an integral operator on $\S^2$ with a smooth kernel
${\rm s}_2(\omega,\omega';\lambda)\in C^\infty(\S^2\times\S^2\times\R^*_+)$.
$S_2(\lambda)$ belongs to the Hilbert-Schmidt class.
\end{prop}
\demo
Let $\psi_0(x,\xi)=\exp(i<\xi,x>)$, then the kernel of the operator $S_2(\lambda)=2i\pi\Gamma_0(\lambda)T_+^*R(\lambda+i0)T_-\Gamma_0^*(\lambda)$ is formaly defined by the expression
\begin{equation}\label{eq:kers2}
{\rm s}_2(\omega,\omega';\lambda)={i\sqrt{\lambda}\over8\pi^2}
\left( T_+^*R(\lambda+i0)T_-\psi_0(\cdot,\sqrt{\lambda}\omega'),
\psi_0(\cdot,\sqrt{\lambda}\omega)\right)_{\LL^2(\R^3)}.
\end{equation}
Formula (\ref{eq:kers2}) is automatically justified if its right hand side is a continuous function of $\omega,\omega',\lambda$. The derivatives $\partial_\omega^\alpha\partial_{\omega'}^{\alpha'}\partial_\lambda^m{\rm s}_2(\omega,\omega';\lambda)$ consist in a sum of terms of the form
\begin{eqnarray*}
\lefteqn{\left( T_+^*R^n(\lambda+i0)T_-\left<x\right>^{\beta'}\psi_0(\cdot,\sqrt{\lambda}\omega'), \left<x\right>^{\beta}\psi_0(\cdot,\sqrt{\lambda}\omega)\right)_{\LL^2(\R^3)}}\\
&=&\left(\left<x\right>^{-n} R^n(\lambda+i0)\left<x\right>^{-n}Q_-\left<x\right>^{-2}\psi_0(\cdot,\sqrt{\lambda}\omega'), Q_+\left<x\right>^{-2}\psi_0(\cdot,\sqrt{\lambda}\omega)\right)_{\LL^2(\R^3)}
\end{eqnarray*}
with $ Q_+=\left<x\right>^nT_+\left<x\right>^{\beta+2},Q_-=\left<x\right>^nT_-\left<x\right>^{\beta'+2}\in{\cal S}^m$ for all $m\in\Z$ and $\vert\alpha\vert+\vert\alpha'\vert+m=\vert\beta\vert+\vert\beta'\vert+n$. Since the operators $Q_\pm$ and $\left<x\right>^{-n}R^n(\lambda+i0)\left<x\right>^{-n}$ are bounded on $\LL^2(\R^3)$ (by theorem \ref{LAP}) and taking into account that $\left<x\right>^{-2}\psi_0(\cdot,\xi)\in\LL^2(\R^3)$ those expressions are correctly defined and bounded. Finally, since $\psi_0(\cdot,\sqrt{\lambda}\omega)$, $\left<x\right>^{-n}R^n(\lambda+i0)\left<x\right>^{-n}$ are continuous in $\lambda$ and $\omega$, we have shown that ${\rm s}_2(\omega,\omega';\lambda)$ is a $C^\infty$-function. In particular $|{\rm s}_2(\omega,\omega';\lambda)|^2$ is bounded and the Hilbert-Schmidt norm of $S_2(\lambda)$ is finite.\fin

Combining propositions \ref{SAB},\ref{S1} and \ref{S2} we obtain
\begin{thm}\label{SM}
Let $H$ be the operator $(\ref{eq:H})$ with potential $(\ref{eq:potA})$, $S(\lambda)$ be the SM for the pair $H,H_0=-\Delta$ and ${\cal W}(\lambda)$ be the operator of multiplication on $\S^2$ by the function ${\rm w}$ defined in $(\ref{eq:wplus})$. Then the operator $S(\lambda)-{\cal W}(\lambda)$ has an infinitly-smooth kernel, in particular it belongs to the Hilbert-Schmidt class.
\end{thm}

We can now prove the two essentials results on spectral properties of the SM for the pair $H,H_0$.
\begin{thm}\label{spec}
Let $H$ be the operator $(\ref{eq:H})$ with potential $(\ref{eq:potA})$ and $S(\lambda)$ be the SM for the pair $H,H_0=-\Delta$, then the essential spectrum of $S(\lambda)$ is given by $(\ref{eq:spec})$.
\end{thm}
\demo
Since ${\cal W}(\lambda)$ is the operator of multiplication by ${\rm w}$, its (continuous) spectrum coincides with the range of the function ${\rm w}$. Since the function $a$ in (\ref{eq:potA}) is positive and taking into account relation (\ref{eq:flux}) the range of the function (\ref{eq:Theta}) equals the interval $[-\phi,\phi]$ and the spectrum of ${\cal W}(\lambda)$ is the image of this interval by the function $\upsilon\mapsto\exp(i\upsilon)$. Finally, since $S(\lambda)-{\cal W}(\lambda)$ is Hilbert-Schmidt, and also compact, thank's to Weyl theorem the essential spectrum of $S(\lambda)$ coincides with the essential spectrum of ${\cal W}(\lambda)$ that is (\ref{eq:spec}).\fin

\begin{thm}\label{scat}
Let $H$ be the operator $(\ref{eq:H})$ with potential $(\ref{eq:potA})$ and $S(\lambda)$ be the SM for the pair $H,H_0=-\Delta$, then the total scattering cross-section $\Sigma_{tot}(\omega_0;\lambda)$ $($defined by $(\ref{eq:diffscat})$ and $(\ref{eq:totalscat}))$ is finite for all incident direction $\omega_0$.
\end{thm}

\demo
The kernel of the principal part ${\cal W}(\lambda)$ is ${\rm w}(\omega)\delta(\omega,\omega')$ where $\delta$ denotes the Dirac distribution on $\S^2$. In particular its support is concentrated on the diagonal $\omega=\omega'$. Off the diagonal the kernel of $S(\lambda)$ reduces to the sum ${\rm s}_1(\omega,\omega';\lambda)+{\rm s}_2(\omega,\omega';\lambda)$. Since ${\rm s}_1$ and ${\rm s}_2$ are infinitly smooth functions the integral (\ref{eq:totalscat}) converge and the total scattering cross-section is finite for all $\omega_0$.
\fin

\begin{rem}
The result of theorem \ref{scat} is conserved under short-range perturbations $(V_0,A_0)$ if we suppose that (\ref{eq:shortR}) is satisfied for some $\rho>3$.
\end{rem}

\section{The Aharonov-Bohm effect in dimension 3}\label{ABp}

Finally, we want to make some remarks on the Aharonov-Bohm effect. Since the 3 dimensionnal example of magnetic field treated here is compactly supported it is somewhat natural to compare our results to those obtained in the 2 dimensionnal case. Let the Aharonov-Bohm Hamiltonian be the operator $H_{AB}=(D-A_{AB}(x))^2$, on $\LL^2(\R^2)$, with magnetic potential
\begin{equation}\label{eq:potAB}
A_{AB}(x)=a(\theta_x){(-x_2,x_1)\over|x|^2},~~|x|\geq R>0,
\end{equation}
where $a\in C^\infty(\R)$ is a $2\pi$-periodic function of the polar angle $\theta_x$ associated to $x=(x_1,x_2)$. The family of potentials satisfying (\ref{eq:potAB}) includes all compactly supported magnetic field in dimension 2. For potential (\ref{eq:potAB}) an analysis, similar to the one made here, was developed in \cite{ROYA2}. With the notations
$$\phi_{AB}=\int_0^{2\pi}a(\vartheta)d\vartheta,~~
f(\theta)=\int_{\theta}^{\theta+\pi}a(\vartheta)d\vartheta$$
it is shown that for $S_{AB}(\lambda)$, the SM associated to the pair $H_{AB},H_0$, we have that
\begin{equation}\label{eq:ABspec}
\sigma_{ess}\left(S_{AB}(\lambda)\right)=\exp{\left(if(\R)\right)}\cup\exp{\left(-i f(\R)\right)}.
\end{equation}
and the differential scattering cross-section admits the asymptotic
\begin{equation}\label{eq:ABdiff}
\Sigma_{diff}(\omega,\omega_0;\lambda)=
{1\over 2\pi\sqrt{\lambda}}{\sin^2(\phi_{AB}/2)\over\sin^2(\theta/2)}+{\cal O}\left({\ln(\theta)\over\theta}\right),
\end{equation}
as $\omega\to\omega_0$, with $|\omega-\omega_0|=2\sin(\theta/2)$. 
The equations (\ref{eq:ABspec}) and (\ref{eq:ABdiff}) generalize the results obtained  by various authors (see \cite{RUI}) for the potential (\ref{eq:potAB}) with $a=C^{ste}(=\phi_{AB}/(2\pi))$ :
$$\sigma(S_{AB}(\lambda))=\sigma_{pp}(S_{AB}(\lambda))=\{e^{i\phi_{AB}/2},e^{-i\phi_{AB}/2}  \},$$
and 
$$\Sigma_{diff}^{AB}(\omega,\omega_0;\lambda)=
{1\over 2\pi\sqrt{\lambda}}{\sin^2(\phi_{AB}/2)\over\sin^2(\theta/2)}.$$
If we compare (\ref{eq:spec}) to (\ref{eq:ABspec}) we remark that both SM have intermediary spectral properties between the general cases of short and long-range potentials where, respectively, the essential spectrum reduce to $\{1\}$ or covers the whole unit circle (see \cite{ROYA}). From (\ref{eq:ABdiff}) we see that in dimension 2 the total scattering cross-section is infinite except if the magnetic flux $\phi_{AB}\in2\pi\Z$, on the contrary for potential (\ref{eq:potA}), in dimension 3, this situation does not appear. \\

In contrast to the 2-dimensionnal case, only few authors have been interested in the 3 dimensionnal case \cite{TAM,BILROB}, so the potential (\ref{eq:potA}) can be regarded as an interesting example. Since the family of potentials satisfying (\ref{eq:potA}) does not contains all compactly supported magnetic fields in dimension 3 we could not exclude the existence of such fields with infinite total cross-section, but it seems that the situation described in this paper is very general. Indeed, let us consider a magnetic potential $A$ obtained from an arbitrary comptacly supported magnetic field that is 
\begin{equation}\label{eq:compact}
{\rm curl} ~A(x)=0\Longleftrightarrow A(x)=\nabla\Phi(x)
\end{equation}
for large $|x|$ and some regular function $\Phi$. It is quite plausible that we can choose $\Phi=\Phi_\pm$, the solutions of the eikonal equation (\ref{eq:eikonal}), as in the case of potential  (\ref{eq:potA}). If this conjecture is verified then we can generalize the scheme developped here to all potentials satisfiying (\ref{eq:compact}). As in section \ref{WO} we can defined the wave operators of lemma \ref{Jpm} which coincide with the usual one and thus are complete.  Similarly the results of section \ref{SMp} would be generalized. Remarking that the results of propositions \ref{S1} and \ref{S2} do not depend on the phases $\Phi_\pm$ and so they holds for the arbitrary potential $A$. Thus all the singularities of the SM are contained in the term ${\cal W}(\lambda)$ defined by (\ref{eq:SAB}). As shown in the proof of proposition \ref{SAB} the singularities of ${\cal W}(\lambda)$ reduce to the Dirac singularity if the function $\Theta(x,\xi)=\Phi_-(x,\xi)-\Phi_+(x,\xi)$ is independent of $x$ (for large $|x|$). This fact will follows from the initial conjecture since 

$$\nabla_x\Theta(x,\xi)=\nabla_x\Phi_-(x,\xi)-\nabla_x\Phi_+(x,\xi)=A(x)-A(x)=0.$$

Then for any compactly supported magnetic field the SM will reduce to multiplication  by the function $\exp(i\Theta(\omega))$, up to a $C^\infty$-kernel operator. So  in contrast to the 2 dimensional case, the total scattering cross section will always be finite in dimension 3.  A final argument for this conjecture can be found in \cite{yafHE}. In this paper Yafaev shown a result  similar to the one conjectured here, but for short range magnetic potentials, that is the high energy limit of the SM  is the operator of multiplication by $\exp(i\int_\R<A(t\omega,\omega)dt)$. 
\bibliographystyle{alpha}
\bibliography{refer}
\end{document}